\documentclass[p4paper,11pt]{article}

\usepackage{subfig}
\usepackage{graphicx}
\usepackage{authblk}
\usepackage{a4wide}

\title{\bf Performance evaluation of the aerogel RICH counter for the Belle II spectrometer using early beam collision data}

\author[1]{M.~Yonenaga}
\author[2,3]{I.~Adachi}
\author[4]{L.~Burmistrov}
\author[4]{F.~Le~Diberder}
\author[5]{T.~Iijima}
\author[6,1]{S.~Iwata}
\author[1]{S.~Kakimoto}
\author[1]{H.~Kakuno}
\author[7]{G.~Karyan}
\author[8]{H.~Kawai}
\author[9]{T.~Kawasaki}
\author[3]{H.~Kindo}
\author[10]{H.~Kitamura}
\author[1]{M.~Kobayashi}
\author[2]{T.~Kohriki}
\author[9,1]{T.~Konno}
\author[11,12]{S.~Korpar}
\author[13,12]{P.~Kri\v{z}an}
\author[1]{T.~Kumita}
\author[1]{K.~Kuze}
\author[2]{Y.~Lai}
\author[12, \footnote{corrently at Institute of High Energy Physics, Vienna, Austria}]{M.~Mrvar}
\author[7]{G.~Nazaryan}
\author[2,3]{S.~Nishida}
\author[2]{M.~Nishimura}
\author[14]{K.~Ogawa}
\author[10]{S.~Ogawa}
\author[12]{R.~Pestotnik}
\author[12, \footnote{currently at Center for Materials and Microsystems, Fondazione Bruno Kessler, Trento, Italy}]{A.~Seljak}
\author[2]{M.~Shoji}
\author[1]{T.~Sumiyoshi}
\author[8]{M.~Tabata}
\author[1]{S.~Tamechika}
\author[14]{Y.~Yusa}
\author[13,12]{L.~\v{S}antelj}

\affil[1]{Department of Physics, Tokyo Metropolitan University, Hachioji, Japan}
\affil[2]{Institute of Particle and Nuclear Studies (IPNS), High Energy Accelerator Research Organization (KEK), Tsukuba, Japan}
\affil[3]{Department of Particles and Nuclear Physics, SOKENDAI (The Graduate University of Advanced Studies), Tsukuba, Japan}
\affil[4]{4Laboratoire de Laccelerateur Lineaire (LAL), Orsay, France}
\affil[5]{Department of Physics, Nagoya University, Nagoya, Japan}
\affil[6]{Electronics and Information Engineering Course, Tokyo Metropolitan College of Industrial Technology, Shinagawa, Japan}
\affil[7]{Experimental Physics Division, Alikhanian National Science Laboratory, Yerevan, Armenia}
\affil[8]{Department of Physics, Chiba University, Chiba, Japan}
\affil[9]{Department of Physics, Kitasato University, Sagamihara, Japan}
\affil[10]{Department of Physics, Toho University, Funabashi, Japan}
\affil[11]{Faculty of Chemistry and Chemical Engineering, University of Maribor, Maribor, Slovenia}
\affil[12]{Experimental Particle Physics Department, Jo\v{z}ef Stefan Institute, Ljubljana, Slovenia}
\affil[13]{Faculty of Mathematics and Physics, University of Ljubljana, Ljubljana, Slovenia}
\affil[14]{Department of Physics, Niigata University, Nishi, Japan}

\date{}

\begin{document}

\maketitle

\begin{abstract}
The Aerogel Ring Imaging Cherenkov (ARICH) counter serves as a particle identification device in the forward end-cap region of the Belle II spectrometer.
It is capable of identifying pions and kaons with momenta up to $4 \, {\rm GeV}/c$ by detecting Cherenkov photons emitted in the silica aerogel radiator.
After the detector alignment and calibration of the probability density function, we evaluate the performance of the ARICH counter using early beam collision data.
Event samples of $D^{\ast +} \to D^0 \pi^+ (D^0 \to K^-\pi^+)$ were used to determine the $\pi(K)$ efficiency and the $K(\pi)$ misidentification probability.
We found that the ARICH counter is capable of separating kaons from pions with an identification efficiency of $93.5 \pm 0.6 \, \%$ at a pion misidentification probability of $10.9 \pm 0.9 \, \%$.
This paper describes the identification method of the counter and the evaluation of the performance during its early operation.
\end{abstract}

\section{Introduction}
The Belle II experiment at the SuperKEKB asymmetric $e^+e^-$ collider~\cite{belle2:1,belle2:2} is searching for new physics beyond the Standard Model by studying rare processes, and aims to collect a high statistics data set corresponding to an integrated luminosity of $50 \, {\rm ab}^{-1}$.
Its predecessor, the Belle experiment~\cite{belle}, was able to identify pions and kaons with momenta up to $2 \, {\rm GeV}/c$ in the forward end-cap region.
However, in the Belle II experiment identification of charged particles with higher momenta up to $4 \, {\rm GeV/}c$ is required to search for new physics in the processes, such as $B \to \rho \gamma (\rho \to \pi\pi)$.
To identify high momentum particles in the forward end-cap region, we employ a new device called the Aerogel Ring Imaging Cherenkov (ARICH) counter.
The ARICH counter construction was finished in October 2018 and the detector was installed into the Belle II spectrometer in December 2018.
At that point, ARICH was integrated into the operation of the Belle II spectrometer.
We evaluate the particle identification performance of ARICH with early beam collision data that were collected between March and June 2019 and correspond to an integrated luminosity of $5.15 \, {\rm fb^{-1}}$.

\section{The ARICH}
\begin{figure}[!h]
\centering
\includegraphics[width=0.8\textwidth]{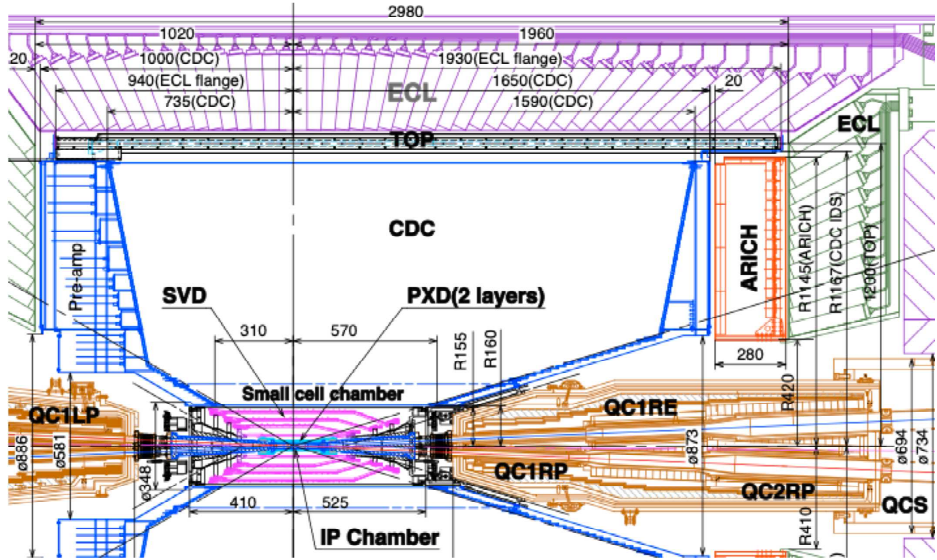}
\caption{Horizontal sectional view of the Belle II spectrometer. The dimensions on the figure are in {\rm mm}.}
\label{b2loc}
\end{figure}

The ARICH counter is located in the forward end-cap of the Belle II spectrometer as shown in Fig.~\ref{b2loc}. This is a novel type of particle identification device that has been developed for the Belle II experiment~\cite{arich:1,arich:2}.
It has a shape of a rectangular toroid with an outside radius of $1145 \, {\rm mm}$, an inside radius of $420 \, {\rm mm}$ and a length of $280 \, {\rm mm}$.
The components along the beam line, starting with the closest to the collision point, are a $40 \, {\rm mm}$ thick radiator, assembled from 248 silica aerogel tiles arranged in two layers, an expansion space of $160 \, {\rm mm}$, and an $80 \, {\rm mm}$ thick photon detector comprised of 420 Hybrid Avalanche Photo Detectors (HAPDs) and readout electronics.

\subsection{Principle}
\begin{figure}[!h]
\centering
\includegraphics[width=0.5\textwidth]{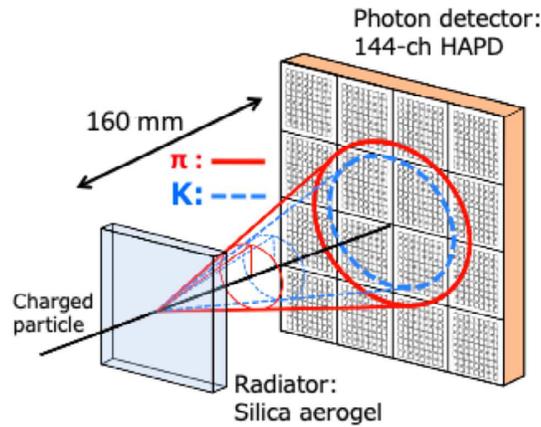}
\caption{The principle of the particle identification of the ARICH counter.}
\label{priari}
\end{figure}

The principle of the particle identification in ARICH is shown in Fig.~\ref{priari}.
ARICH uses Cherenkov photons emitted in the silica aerogel radiator when the particle velocity exceeds the speed of light in the medium.
Emitted photons are detected by the HAPDs and the emission angles of Cherenkov photons are measured.
As usual in a RICH counter, by measuring the angle of Cherenkov photons, the mass of the particle can be determined through the relation
\begin{equation}
\cos{\theta} = \frac{\sqrt{(mc/p)^2+1}}{n},
\end{equation}
where $\theta$ is the emission angle of Cherenkov photons, $m$ is the mass of the particle, $n$ is the refractive index of the silica aerogel, and $p$ is the momentum of the particle as measured by the tracking system.

\subsection{Components of the detector}
The main components of the ARICH counter, the silica aerogel radiators, the HAPD photo-sensors and the readout electronics, were all developed for the Belle II experiment.

\subsubsection{Silica aerogel radiator}
\begin{figure}[!h]
\centering
\includegraphics[width=0.5\textwidth]{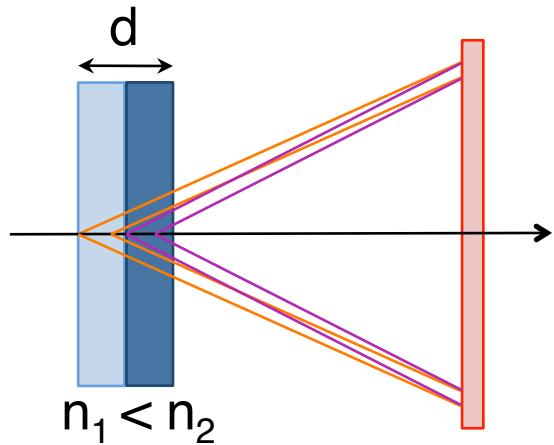}
\caption{The conceptual diagram of the dual layer scheme.}
\label{aerogel}
\end{figure}

The ARICH counter uses large-area silica aerogel tiles with a long transmission length~\cite{aero1,aero2,aero3}.
A dual-layer scheme is adopted to increase the number of detected Cherenkov photons without degrading the angular resolution~\cite{dualaero:1,dualaero:2,dualaero:3}.
By optimizing the difference of refractive indices of the two  layers, Cherenkov photons emitted from both layers overlap, as illustrated in Fig.~\ref{aerogel}, resulting in a narrower ring image.
We choose refractive indices 1.045 and 1.055 for the upstream and the downstream aerogel layer, respectively.
The threshold momenta to emit Cherenkov photons in the radiator with a refractive index of 1.055 are $0.42 \, {\rm GeV}/c$ for pions and $1.47 \, {\rm GeV}/c$ for kaons.
The silica aerogel radiator is composed of 124 pairs of wedge-shaped tiles arranged in four concentric rings with 22, 28, 34, and 40 pairs.
The tiles were cut out from $180 \times180 \times 20 \, {\rm mm^3}$ aerogel blocks.
The refractive indices and transmission lengths of produced silica aerogel tiles are summarized in Table~\ref{tab:aero}.
The upstream and downstream aerogel tiles were paired in a way to keep the difference of their refractive indices within the required range $0.010 \pm 0.002$, which is optimal for the Cherenkov rings to overlap. For the installed tile pairs this difference ranges from 0.0095 to 0.0104.

\begin{table}[]
\centering
\caption{The average of the refractive indices and transmission lengths of produced silica aerogel tiles. The uncertainties show the standard deviations of the distributions.}
\begin{tabular}{lcc}\hline
Target refractive index				& 1.045					& 1.055 \\ \hline
Refractive index at $405 \, {\rm nm}$		& $1.0451 \pm 0.0007$		& $1.0547 \pm 0.0007$ \\
Transmission length at $400 \, {\rm nm}$	& $47.3 \pm 3.1 \, {\rm mm}$	& $36.0 \pm 2.7 \, {\rm mm}$\\ \hline
\end{tabular}
\label{tab:aero}
\end{table}

\subsubsection{HAPD photo-sensors}
\begin{figure}[!h]
\centering
\includegraphics[width=0.5\textwidth]{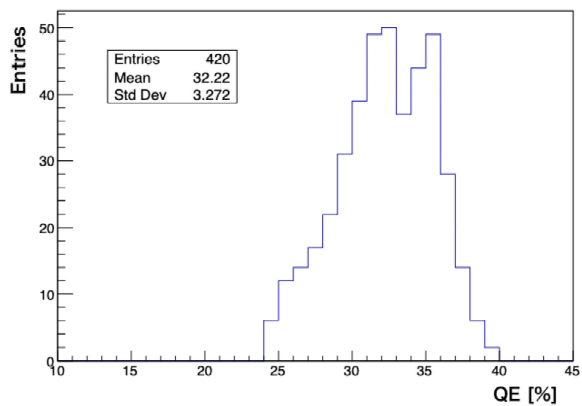}
\caption{Distribution of quantum efficiency of installed HAPDs.}
\label{qe}
\end{figure}

The photon detector of ARICH is required to have the following properties: single-photon detection capability with the pad size smaller than $5 \, {\rm mm}$, tolerance to the high magnetic field ($B = 1.5 \, {\rm T}$), and a sufficient radiation hardness to last throughout the 10 years' operation of Belle II (radiation dose $100 \, {\rm Gy}$ and neutron fluence equivalent to $10^{12} \, 1\, {\rm MeV}$ neutrons per ${\rm cm^2}$).
The HAPDs which have been developed with Hamamatsu Photonics K.K. fulfill the requirements as the photon detector~\cite{hapd,hapdperf1,hapdperf2}.
The properties of the HAPDs are summarized in Table~\ref{tab:hapd:spec}.
The HAPDs are constructed of a quartz window with a photocathode, a vacuum tube and four Avalanche Photo Diode (APD) chips.
Each APD chip is segmented into $6 \times 6$ pixels of an area of $4.9 \times 4.9 \, {\rm mm^2}$, totaling to 144 channels for the full HAPD.
The photon detection with the HAPD proceeds in two steps: bombardment of the APD by a photoelectron and the avalanche multiplication in the APD chip.
The photoelectrons are accelerated in the high electric field by applying a negative high voltage to the photocathode; they produce about 1800 electron-hole pairs in the APD.
Each electron is further multiplied and produces 40 electron-hole pairs in the high field region formed by the reverse bias voltage applied to the APD.
As a result, the total gain of the HAPD is about 72000.
HAPDs are arranged in seven concentric rings with 42, 48, 54, 60, 66, 72, and 78 detectors, 420 HAPDs in total.
Fig.~\ref{qe} shows the distribution of the average quantum efficiency over the area of the photocathode at 400 nm of all  installed HAPDs.
The mean value of the quantum efficiency of the installed HAPDs is $32.2 \, \%$.
They were distributed randomly on the photon detector plane to minimize the position dependence of detection efficiency for the Cherenkov light.
HAPDs are performing as expected during the operation of the Belle II spectrometer~\cite{hapdperf3}.

\begin{center}
\begin{table}[]
\centering
\caption{The specification of the HAPD.}
\begin{tabular}{lc}\hline
Size					& $73 \times 73 \times 28 \, {\rm mm^3}$ \\
Number of channels		& $12 \times 12 = 144 \, {\rm ch}$ \\
Channel size			& $4.9 \times 4.9 \, {\rm mm^2}$ \\
Effective area			& $65 \, \%$ \\
Photo-cathode material	& Bialkali \\
Quantum efficiency		& $\sim 28 \, \%$ at $400 \, {\rm nm}$\\
Bombardment gain		& $\sim$1800 \\
Avalanche gain			& $\sim$40 \\
Total gain				& 72000 \\
Capacitance			& $80 \, {\rm pF}$\\ \hline
\end{tabular}
\label{tab:hapd:spec}
\end{table}
\end{center}

\subsubsection{Readout electronics}
Readout electronics consist of Front End Boards (FEBs) and Merger Boards (MBs)~\cite{ro1,ro2,Nishida:2012tua}.
The FEB is directly attached to the HAPD and has four Application Specific Integrated Circuit (ASIC) chips developed for the HAPD readout~\cite{asic} and a Xilinx Spartan-6 field programmable gate array (FPGA).
The ASIC chip continuously amplifies, shapes and discriminates 36 signals coming from a quarter of the HAPD.
FPGA continuously samples ASIC digital outputs and stores all hits into a pipeline with a clock cycle of 126 ns. To measure the ring image, the digital hit information within the time window of 504 ns (4 bits) is sent to a MB after a trigger is received\footnote{Since the expected interaction rate at the design luminosity of SuperKEKB is $\sim 60 \, {\rm kHz}$, this time window is sufficiently short to suppress the mixing of the hits from different interactions.}.
Up to six FEBs are connected to one MB; each MB is equipped with a Xilinx Virtex-5 FPGA.
It serializes the data from FEBs and suppresses the data of channels that have not been hit to reduce the data size.
The MB sends merged data to the back-end Belle II central data acquisition system.
Each MB has two RJ45 connections, for trigger and JTAG(Joint Test Action Group), and an optical fiber for data flow; configuration of the FEBs is set via the MBs.
We use 420 FEBs with 60480 channels in total, and 72 MBs.
All components were tested to withstand the radiation dose 100 Gy and neutron fluence equivalent to $10^{12} \, 1\, {\rm MeV}$ neutrons per ${\rm cm^2}$ without permanent damage, but occasional reconfiguration of FPGAs will be required due to single-event upsets \cite{Higuchi:2012zz}.

\section{Particle identification method in ARICH}
Particle identification by the ARICH counter is based on the comparison between the observed pattern of photons and the probability density function (PDF), which describes the expected distribution of Cherenkov photons and background hits on the photon-detector plane for given charged track parameters and the assumed particle type hypothesis.
For each charged track passing through ARICH, we evaluate the value of the likelihood function for six particle hypotheses: electron, muon, pion, kaon, proton, and deuteron.
The likelihood function for a particle hypothesis $h$ is defined as 
\begin{equation}
\mathcal{L}_h = \prod_{\textrm{all pixels}}{p_{h,i} (m_{h,i})}; ~~~~~~ p_{h,i}(m_{h,i}) = \frac{e^{-n_{h,i}} n^{m_{h,i}}_{h,i}}{m_{h,i}!},
\end{equation}
where the product runs over all pixels of the whole ARICH counter, and $p_{h,i} (m_{h,i})$ is the probability of observing $m_{h,i}$ hits on the $i$ th pixel, while $n_{h,i}$ hits are being expected on average (for an assumed hypothesis $h$)~\cite{cal,pid1,pid2,pid3}.
Since we do not discriminate multiple-photon from single-photon hits in ARICH, $p_{h,i}(m_i)$ can be rephrased as 
\begin{eqnarray}
p_{h,i} (\textrm{no hit})&=& e^{-n_{h,i}}\\
p_{h,i} (\textrm{hit})&=& 1 - p_{h,i}(\textrm{no hit}) = 1 - e^{-n_{h,i}}.
\end{eqnarray}
Using Eq. (3) and Eq. (4), Eq. (2) can be rewritten as 
\begin{equation}
\ln{\mathcal{L}_h} = - N_h + \sum_{\textrm{hit $i$}} [n_{h,i} + \ln{(1-e^{-n_{h,i}})}],
\end{equation}
where $N_h$ is the expected total number of hits, and the sum runs only over pixels that were hit\footnote{This greatly simplifies the evaluation of Eq. (2) since the expected number of hits only needs to be calculated for pixels that were hit, and not for all of them.}.
$N_h$ is calculated from the expected number of emitted photons, reduced for scattering and absorption in the silica aerogel, the geometrical acceptance and detection efficiency, and the expected number of background hits. The value of 
$n_{h,i}$ is obtained by integrating the PDF for the assumed particle hypothesis $h$ over the surface of the pixel $i$.
To evaluate the likelihood, only the photon-detector area corresponding to the reconstructed Cherenkov angle between $0.1 \, {\rm rad}$ and $0.5 \, {\rm rad}$ is taken into account.
The reconstructed Cherenkov angle is calculated as the angle between the momentum vector associated with the track and the initial direction of the photon emitted from the mean emission point in the aerogel and detected at a given pixel on the photon detector.
We describe the construction of PDF in more detail in the next section.
The separation between kaons and pions is performed by imposing selection criteria on the likelihood ratios which are defined as
\begin{eqnarray}
R_{K/\pi} &=& \frac{\mathcal{L}_K}{\mathcal{L}_K + \mathcal{L}_\pi}\\
R_{\pi/K} &=& \frac{\mathcal{L}_\pi}{\mathcal{L}_K + \mathcal{L}_\pi} = 1 - R_{K/\pi}.
\end{eqnarray}

\subsection{PDF construction}
The PDF is constructed from track-correlated and -uncorrelated parts. First the track-correlated part is constructed as a function of reconstructed Cherenkov angle. This angle corresponds to the Cherenkov angle for unscattered Cherenkov photons.
The number of expected hits on a given channel ($n_{h,i}$) is then obtained by projecting the correlated part of the PDF onto the photon-detector plane and integrating it over the surface of the pixel $i$, and adding the uncorrelated part of the PDF.
The PDF is constructed from the following components.

\begin{itemize}
\setlength{\leftskip}{-0.7cm}

\item[] (1) Cherenkov photons emitted in silica aerogel.
\begin{itemize}
\item[]
A particle passing through the ARICH will emit Cherenkov light in the silica aerogel if its velocity exceeds the speed of light in aerogel.
The photons originated from this process can be described by two Gaussian peaks, which have the same center at the expected Cherenkov angle, in the Cherenkov angle distribution.
The narrow one corresponds to unscattered Cherenkov photons and the broad one to Rayleigh-scattered photons in silica aerogel, and photons passing through the photo-cathode and detected after being reflected from the APD surface (the fraction of this broader peak is about 5\%).
The magnitudes of both peaks are proportional to the expected number of emitted photons in the aerogel and the widths and fraction of each are determined by fitting the reconstructed Cherenkov angle distributions in control samples, as described in the next paragraph.
\end{itemize}

\item[] (2) Track correlated background sources
\begin{itemize}
\item[]
Cherenkov photons are also emitted when a particle passes through the quartz window of an HAPD.
They can be converted to photoelectrons on their first impact on the photocathode or after repeated total internal reflections in the quartz window and contribute to PDF at small Cherenkov angles.
Part of the background comes also from Cherenkov photons emitted by track-related delta rays. 
This component is separately determined for the cases when the particle enters the quartz window or not, as discussed below.
\end{itemize}

\item[] (3) Random background sources
\begin{itemize}
\item[]
Hits due to electronics noise, HAPD dark counts, Cherenkov photons from other tracks, beam background etc., contribute to random background.
This part is not correlated to the track and we model it by uniform distribution over the photon detector area.
The level is determined from the average number of the off-time hits per channel per event.
\end{itemize}

\end{itemize}
We determine each component by tuning the expected distribution of hits as a function of the reconstructed Cherenkov angle; the details of the procedure are described in following sections.

\subsection{PDF calibration}
For efficient particle identification, the PDF needs to be calibrated to accurately represent the expected distribution of hits for each particle hypothesis.
The calibration is performed using real data samples by tuning the expected distribution of reconstructed Cherenkov angles to the measured distribution.
The expected distribution is constructed as a sum of the following items.
\begin{description}
	\item[Sum of two Gaussian functions.]\mbox{}\\
This part represents the contribution of the Cherenkov photons emitted in silica aerogel (1).
The same parameter is user for the mean values of both Gaussian functions.
The width of the narrower Gaussian function is fixed to the value explained in Sect.~\ref{sec:reso}. The remaining four parameters are free.
	\item[The sum of exponential and first- or second-order polynomial.]\mbox{}\\
This function is used to describe the track correlated background (2). The first-order polynomial is used for the case when a particle misses quartz window and the second order for when the window was hit.
All parameters are treated as free.
	\item[The constant part.]\mbox{}\\
Constant contribution from random hits described in Sect.~\ref{sec:flat}. 
\end{description}
\subsubsection{Samples}
\begin{figure}[!h]
\centering\includegraphics[width=0.5\textwidth]{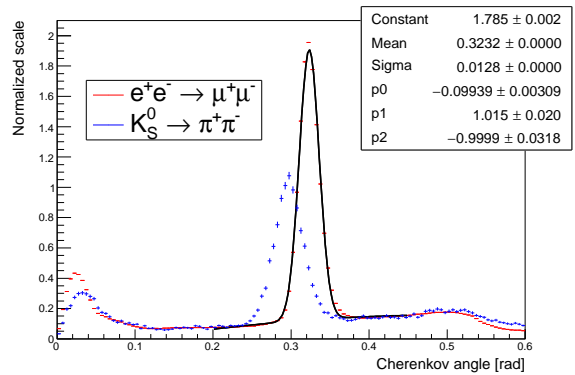}
\caption{Distributions of the reconstructed Cherenkov angles for the $e^+e^- \to \mu^+\mu^-$ sample with muon momentum selection $6.4 \, {\rm GeV}/c < p < 7.0 \, {\rm GeV}/c$, and for the $K_S^0 \to \pi^+\pi^-$ sample with pion momentum selection $1.0 \, {\rm GeV}/c < p < 1.1 \, {\rm GeV}/c$. Both are normalized by the number of tracks. The average number of Cherenkov photons detected in the ring of a high momentum muon is 11.4, as determined by the Gaussian fit (black line), and agrees well with the MC.}
\label{angle}
\end{figure}

The $e^+e^- \to \mu^+\mu^-$ and $K^0_S \to \pi^+\pi^-$ samples which correspond to an integrated luminosity of $2.62 \, {\rm fb^{-1}}$ are used for the PDF calibration.
The $e^+e^- \to \mu^+\mu^-$ events provide a clean sample since this process can be easily identified by requiring exactly two tracks in the event, both identified as muons to reject Bhabha events.
The tracks of this sample only cover a high momentum range, while the $K^0_S \to \pi^+\pi^-$ sample is used to calibrate the low momentum range.
The reconstructed Cherenkov angle distributions for both samples for selected momentum ranges are shown in Fig.~\ref{angle}.
The number of observed photons are obtained by fitting with a Gaussian and a second-order polynomial in case of the  $e^+e^- \to \mu^+\mu^-$ samples.
\subsubsection{The width of the Cherenkov angle distribution}\label{sec:reso}
\begin{figure}[!h]
\centering
\includegraphics[width=0.5\textwidth]{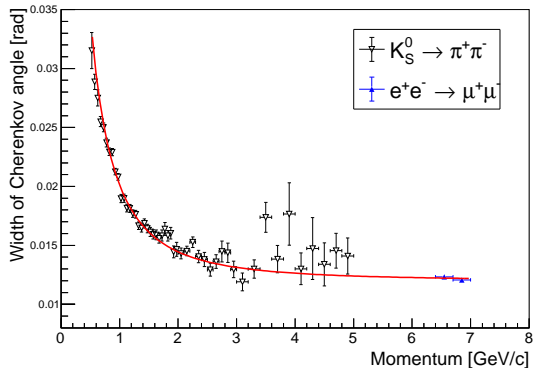}
\caption{Width of the narrow peak in the measured Cherenkov angle distribution as a function of momentum of the particle; the red curve is the fitted function, as discussed in the text.}
\label{reso}
\end{figure}

The width of the measured Cherenkov angle distribution, which is the standard deviation of the main narrow Gaussian, depends on the tracking resolution, which strongly depends on the momentum.
Therefore, we implement the width of Cherenkov angle as a function of the particle momentum.
To measure the width of Cherenkov angle as a function of momentum, we use a $e^+e^- \to \mu^+\mu^-$ sample for momenta around $ 7\, {\rm GeV}/c$ and the $K_S^0 \to \pi^+\pi^-$ sample for momenta below $5 \, {\rm GeV}/c$.
The width of the measured Cherenkov angle distribution as a function of momentum is shown in Fig.~\ref{reso}.
In addition to the PDF-related part $\sigma_{\rm PDF}(p)$, the measured width also contains a contribution from the photon-detector position resolution, i.e., the uncertainty due to pad size, which is estimated to be $\sigma_{\rm pad} = 6.4 \, {\rm mrad}$ and is included in the fit function as a constant part.
We obtain a function of width of Cherenkov angle by fitting the measured distribution with a function
\begin{equation}
f(p) = \sigma_{\rm PDF}(p) \oplus \sigma_{\rm pad};\quad
\sigma_{\rm PDF}(p) =  \sqrt{(A/p)^2+B^2},\end{equation}
motivated by the momentum-dependency of angular dispersion due to multiple scattering, where $A$ and $B$ are parameters determined by the fitting.
\subsubsection{Random hits}\label{sec:flat}
\begin{figure}[!h]
\centering
\includegraphics[width=0.5\textwidth]{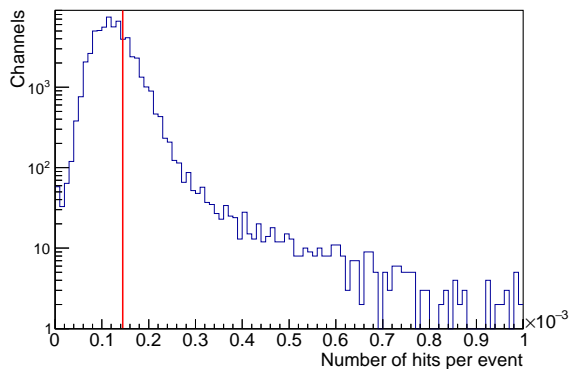}
\caption{Distribution of the average number of hits per event for all channels. The red line is the average value which is implemented in the PDF.}
\label{flat}
\end{figure}

For a single event, hits are recorded in the time window of $504 \, {\rm ns}$ divided into four equal bins.
Correlated hits are recorded in the middle two bins and from the others we can estimate the rate of the uncorrelated background.
Since the contribution of random hits to the overall background is relatively small, we assume the rate of those to be the same for all pixels.
Fig.~\ref{flat} shows the distribution of the average number of random hits per event for all detector channels.
The average number of random hits per event and per pixel was calculated to be $1.4 \times 10^{-4}$; its value is also indicated on the figure.

\subsubsection{Comparison}
\begin{figure}[!h]
\centering
\includegraphics[width=0.5\textwidth]{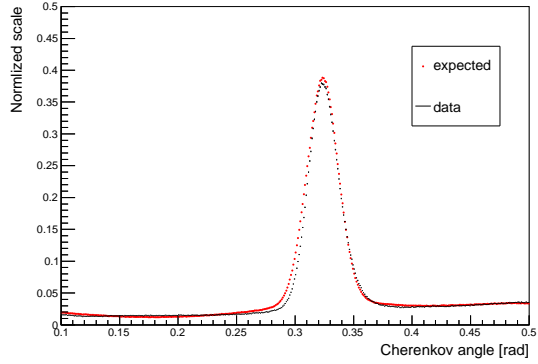}
\caption{Comparison of Cherenkov angle distributions for the data and for the PDF expectation using the $e^+e^- \to \mu^+\mu^-$ sample. Both distributions are normalized by the number of tracks.}
\label{mumu}
\end{figure}

\begin{figure}[!h]
\centering
\includegraphics[width=\textwidth]{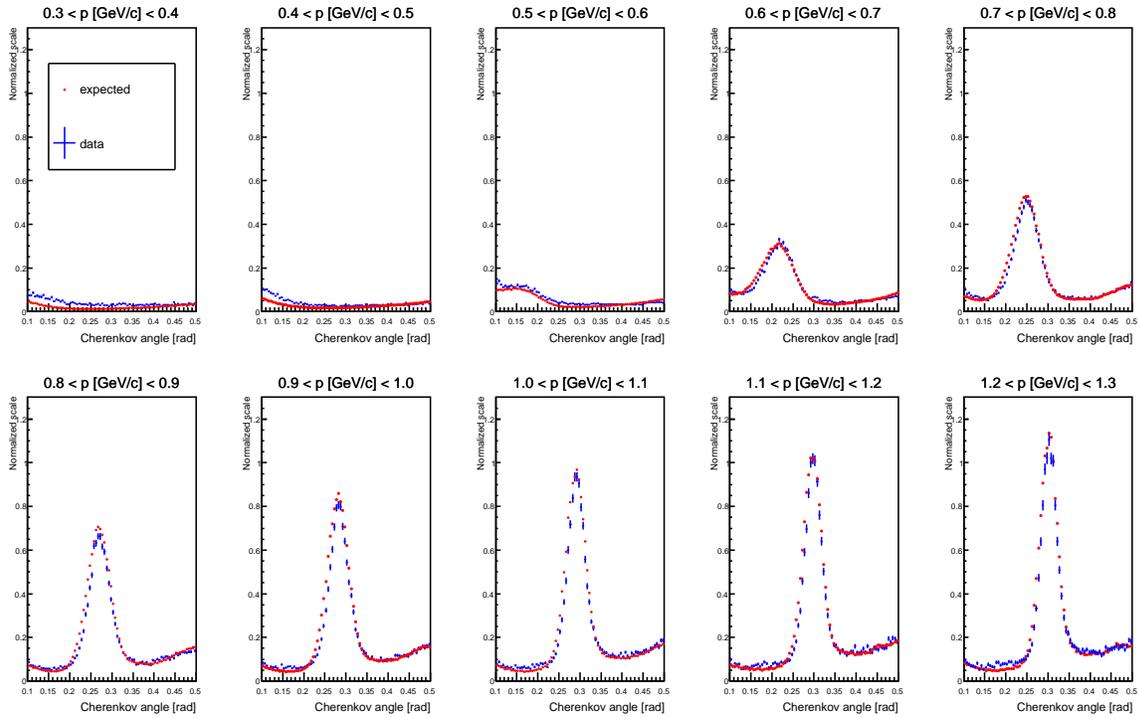}
\caption{Comparison of Cherenkov angle distributions for the data and for the PDF estimation using the $K^0_S \to \pi^+\pi^-$ samples. The blue marker is used for data and the red marker for the PDF estimation. Legend is common for all plots. Both distributions are normalized by the number of tracks. Each plot shows a different momentum range.}
\label{ks0}
\end{figure}

The rest of the PDF parameters are tuned by comparing the reconstructed Cherenkov angle distribution for the measured data samples with the expected distribution obtained by filling the histogram of reconstructed Cherenkov angles of all pixels weighted by the expected average number of hits for the pixel, as evaluated from the PDF.
The results are shown in Figs.~\ref{mumu} and ~\ref{ks0} using the $e^+e^- \to \mu^+\mu^-$  and $K^0_S \to \pi^+\pi^-$ samples.
The results show that the tuned PDF agrees well with the data over a wide range of the momenta.

\section{Performance evaluation}

\subsection{Control sample}
We use $D^{\ast}$ samples from $e^+e^- \to B\bar{B}$ and $e^+e^- \to q\bar{q} \, (q=u,d,s~{\rm and}~c)$ events which correspond to an integrated luminosity of $5.15 \, {\rm fb^{-1}}$, and Monte Carlo (MC) simulation data generated with the present Belle II geometry, which corresponds to an integrated luminosity of $10 \, {\rm fb^{-1}}$.
We use the $D^{\ast +} \to D^0 \pi^+ (D^0 \to K^-\pi^+)$ decays\footnote{The charge-conjugated mode is always implied.} to evaluate the particle identification performance.
This decay can be reconstructed with a relatively low background level without requiring particle identification information for pion and kaon tracks from $D^0$ decays.
A track coming from a $D^0$ decay can be identified by its charge as pion (kaon) if the track has the same (opposite) charge as the ``slow pion'' from the $D^{\ast +}$ decay.
To reconstruct $D^{\ast +}$, we require $ |M_{D^{\ast +}} - M_{D^0} - 0.1454 \, {\rm GeV}/c^2| < 0.0015 \, {\rm GeV}/c^2 $.
To evaluate the performance of ARICH, a $\pi^+$ or $K^-$ candidate from $D^0$ decay is required to enter the ARICH counter.
In addition, more than one hit in the central drift chamber is required.

\subsection{Particle identification performance}
\begin{figure}[!h]
\centering
\subfloat[][$\pi$ tracks]{\includegraphics[width=0.5\textwidth]{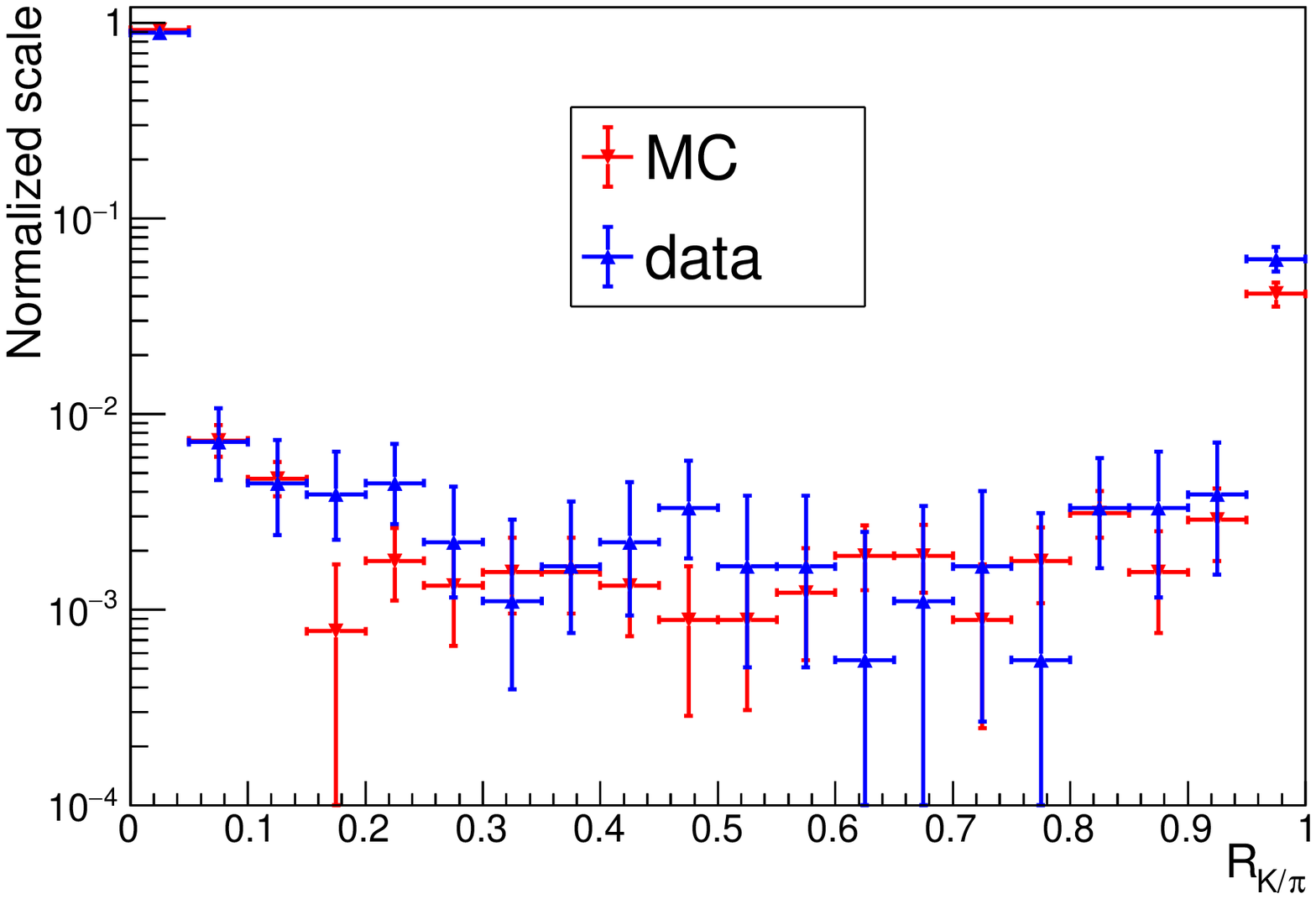}\label{pidpi}}
\subfloat[][$K$ tracks]{\includegraphics[width=0.5\textwidth]{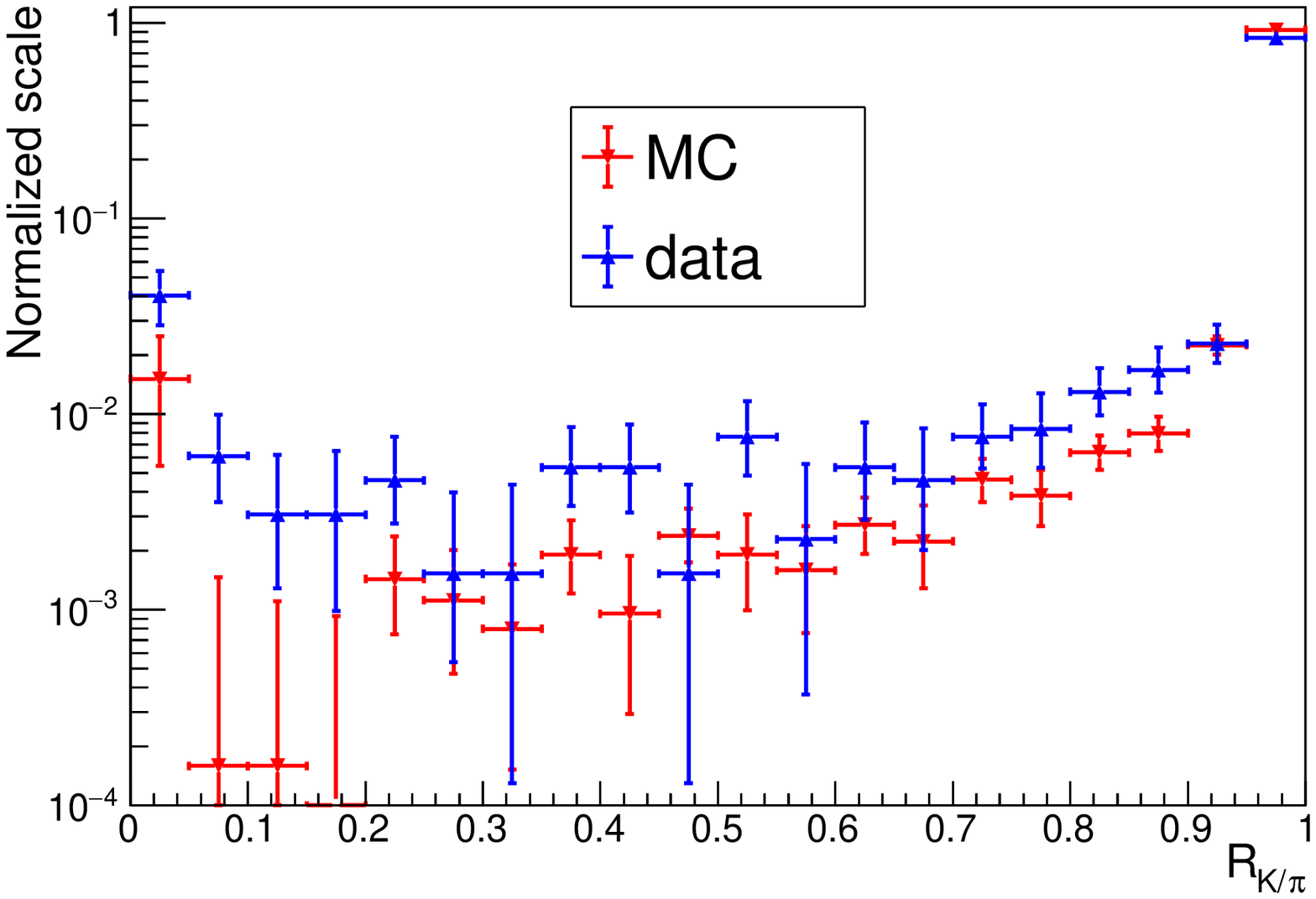}\label{pidk}}
\caption[]{Comparison of the distributions of $R_{K/\pi}$ from MC and measured data for \subref{pidpi} $\pi$ candidate and \subref{pidk} $K$ candidate tracks. All distributions are normalized by the number of tracks.}
\label{pid}
\end{figure}

\begin{figure}[!h]
\centering\includegraphics[width=\textwidth]{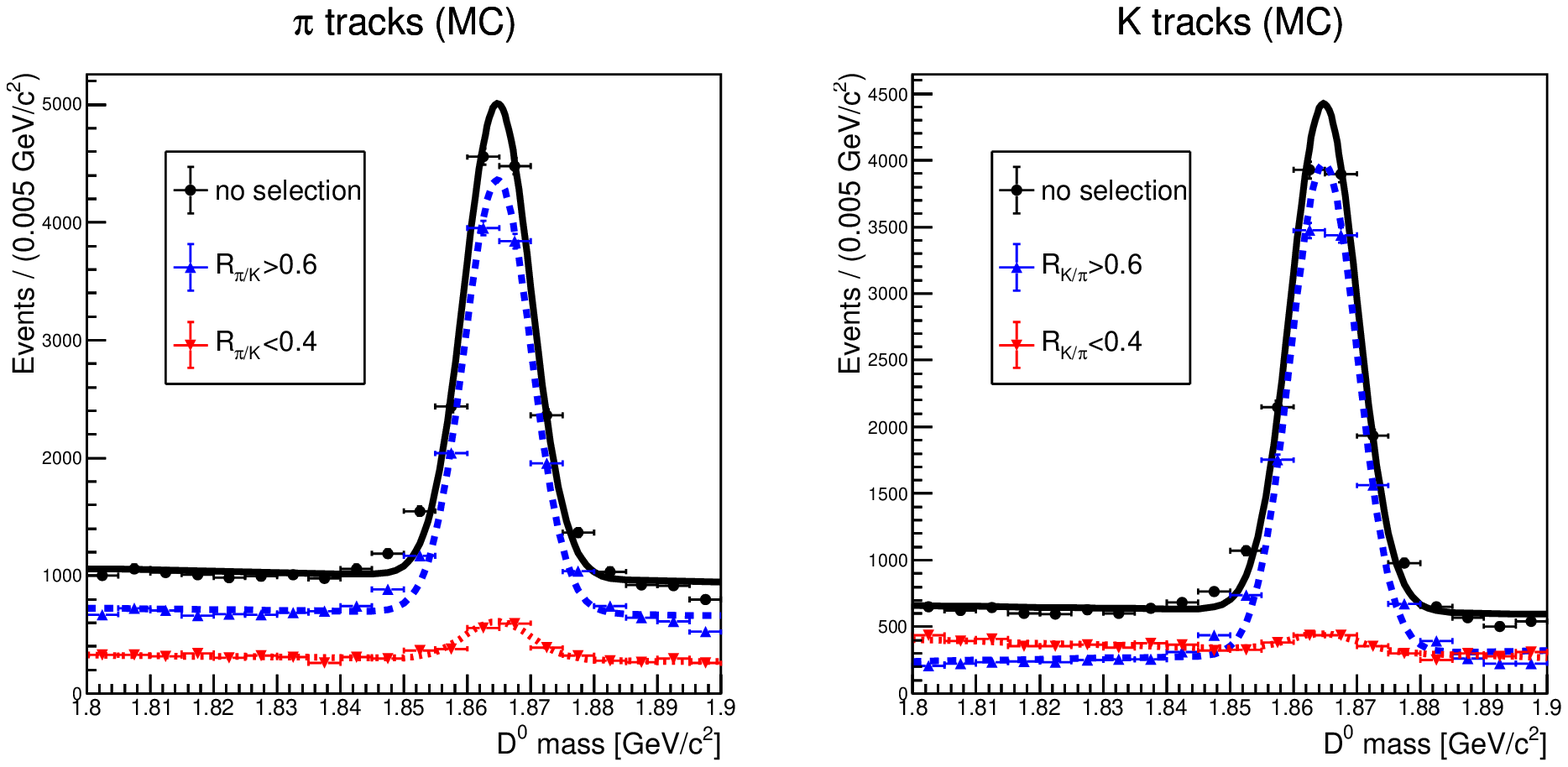}
\caption{Distributions of $D^0$ mass using MC before and after applying $R_{\pi/K} (R_{K/\pi})$ selection criteria. Black solid, blue dashed and red dotted curves show results of the fit for each $D^0$ mass distribution.}
\label{d0mass:mc}
\end{figure}

\begin{figure}[!h]
\centering\includegraphics[width=\textwidth]{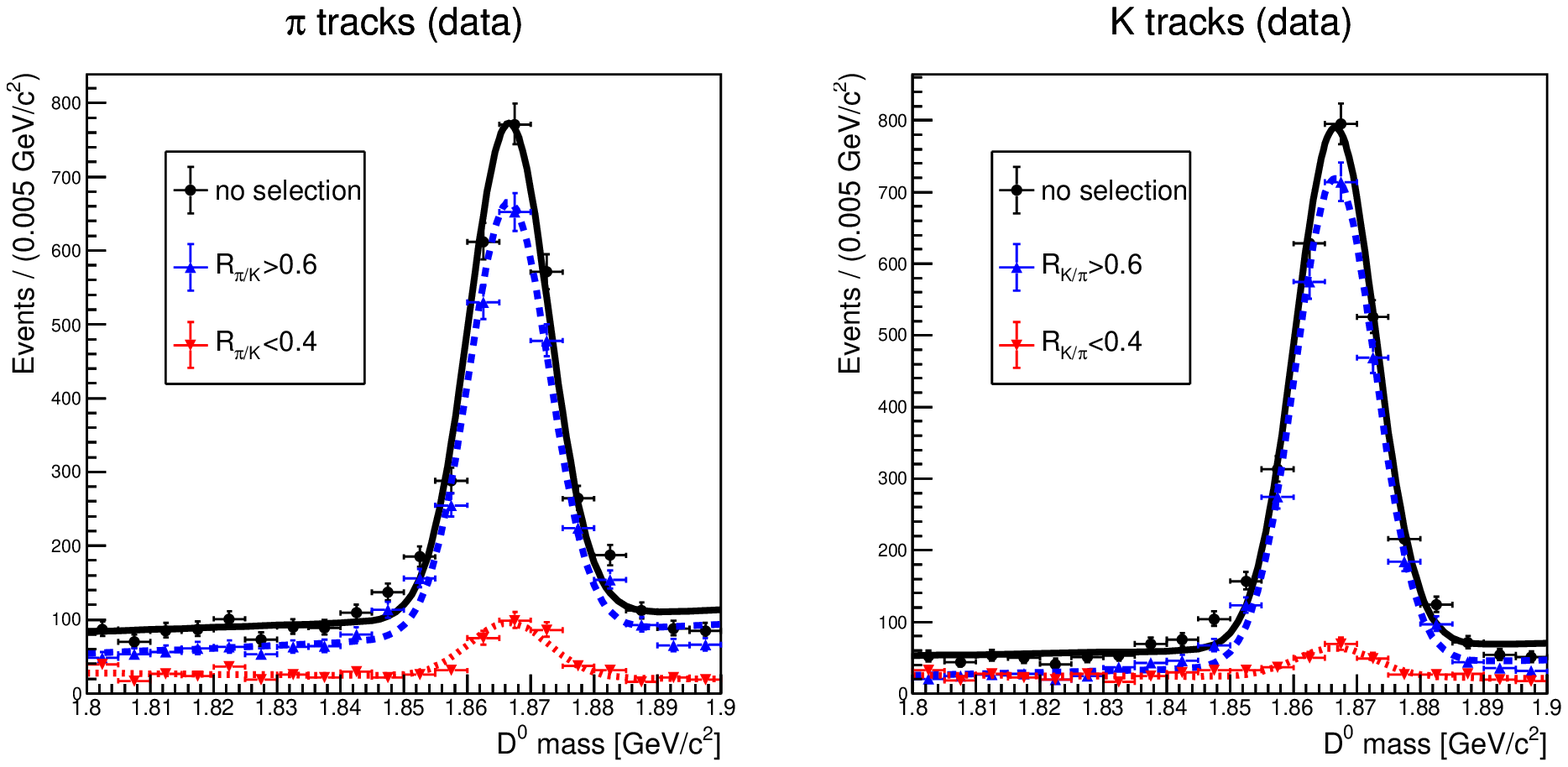}
\caption{Distributions of $D^0$ mass using measured data before and after applying $R_{\pi/K} (R_{K/\pi})$ selection criteria. Black solid, blue dashed and red dotted curves show results of the fit for each $D^0$ mass distribution.}
\label{d0mass:data}
\end{figure}

\begin{figure}[!h]
\centering\includegraphics[width=\textwidth]{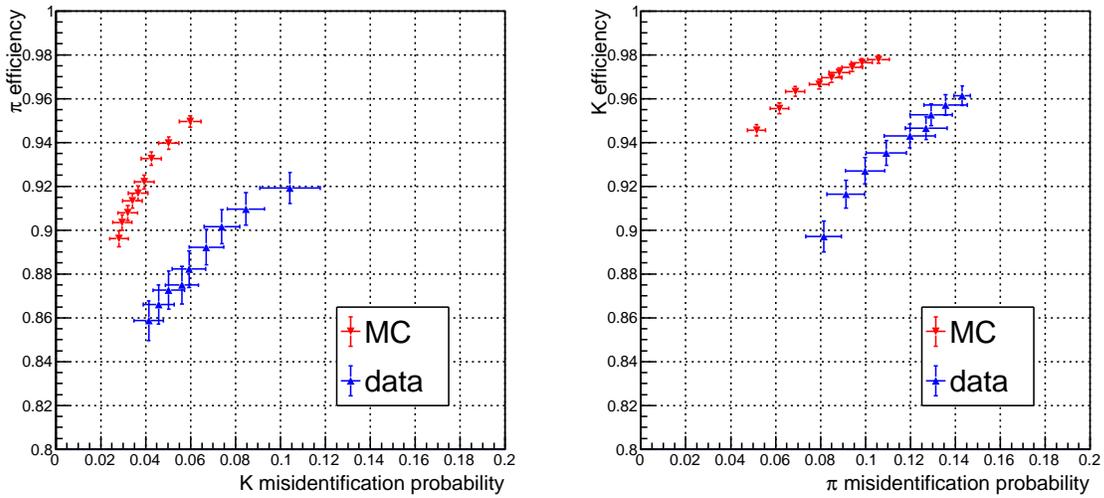}
\caption{ROC curve for the $\pi$ efficiency versus the $K$ misidentification probability (left) and the $K$ efficiency versus the $\pi$ misidentification probability (right).}
\label{roc}
\end{figure}

\begin{figure}[!h]
\centering\includegraphics[width=\textwidth]{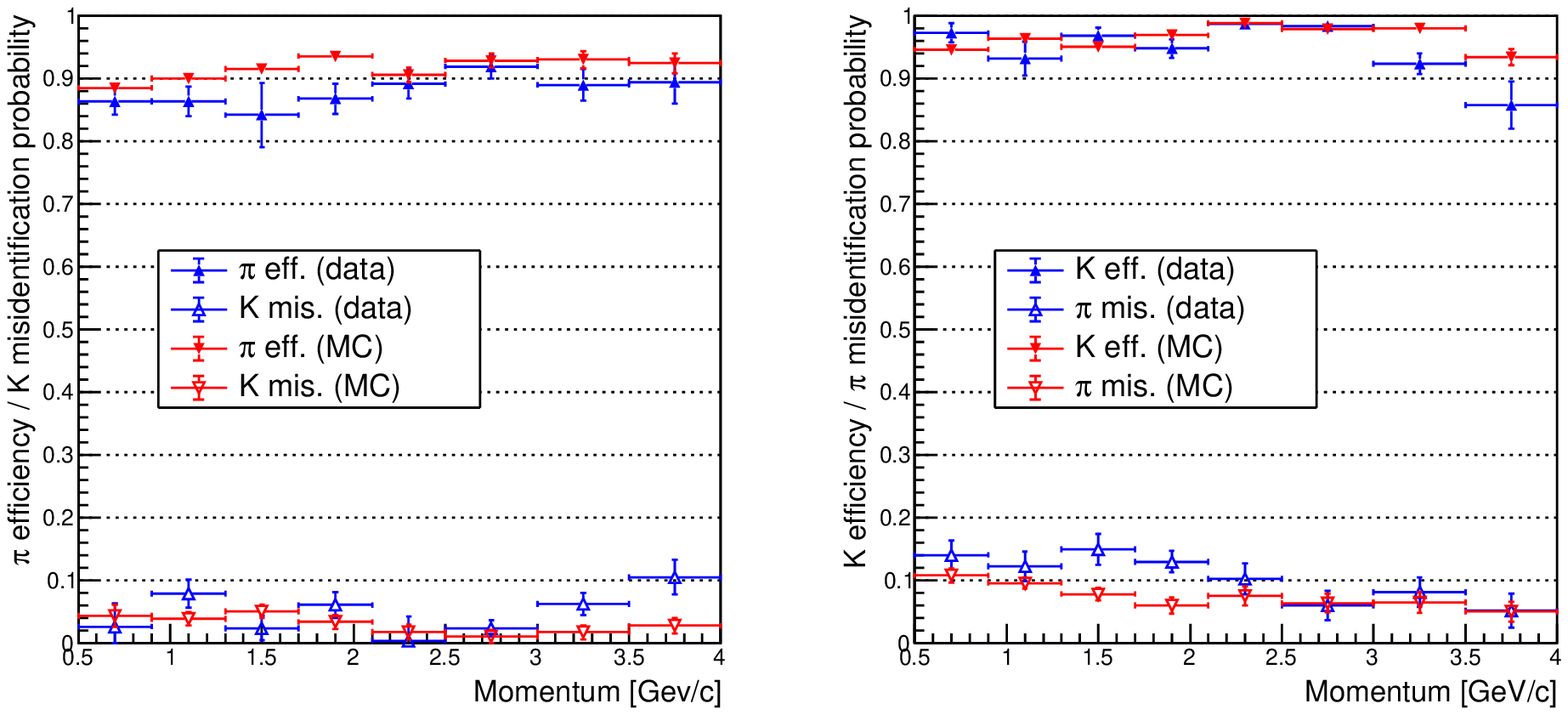}
\caption{ Efficiency and misidentification probability as a function of the momentum: $\pi$ efficiency and $K$ misidentification probability (left), and $K$ efficiency and $\pi$ misidentification probability (right).}
\label{mom}
\end{figure}

\begin{figure}[!h]
\centering\includegraphics[width=\textwidth]{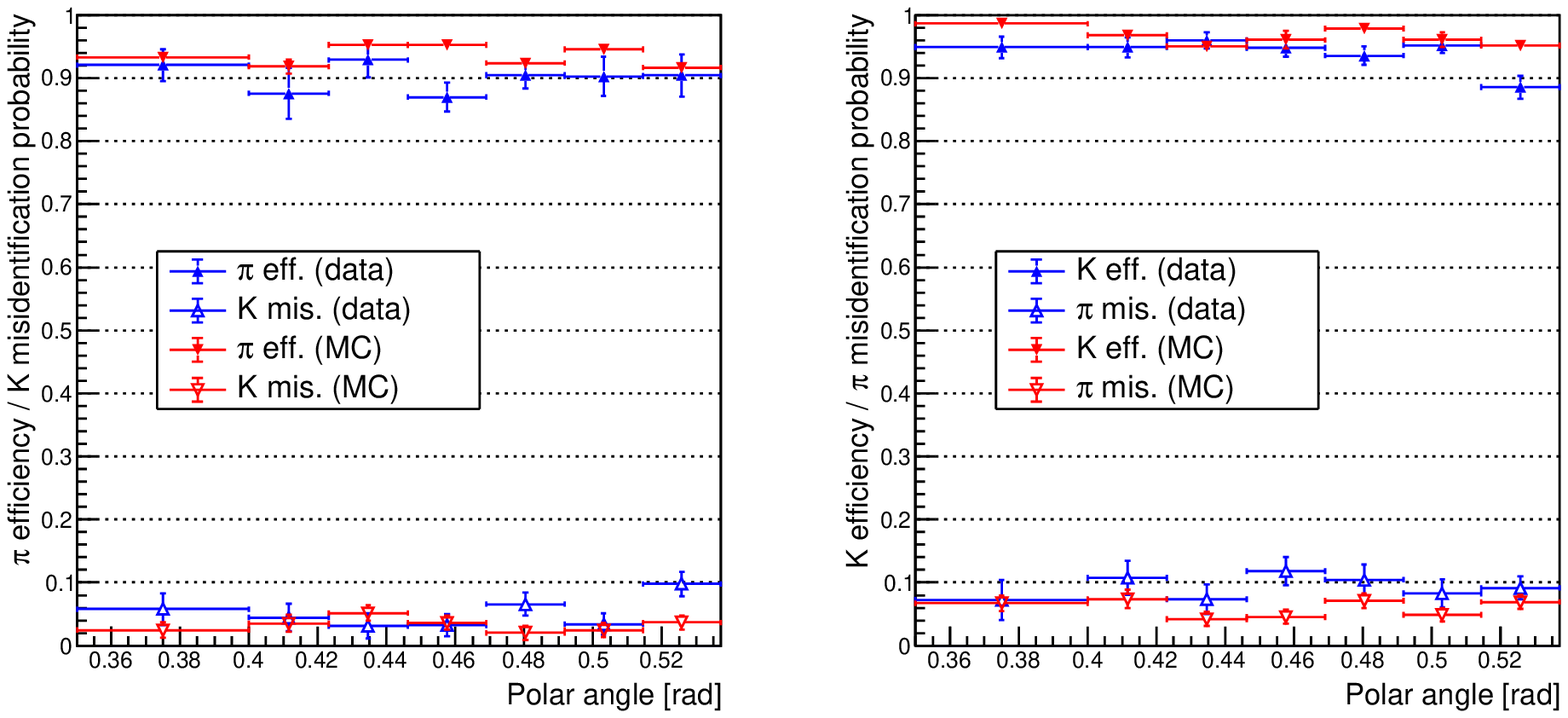}
\caption{Efficiency and misidentification probability as a function of the polar angle: $\pi$ efficiency and $K$ misidentification probability (left), and $K$ efficiency and $\pi$ misidentification probability (right).}
\label{theta}
\end{figure}

The identification performance is discussed in terms of the identification efficiency and the misidentification probability.
Both are determined as a ratio of the number of reconstructed $D^{\ast}$ decays obtained with and without the identification selection criteria applied, i.e.:
\begin{eqnarray}
K \, \textrm{efficiency}&=&\frac{\textrm{number of {\it K} tracks after $R_{K/\pi}>R_{{\rm cut}}$ }}{\textrm{number of {\it K} tracks}}\\
\pi \, \textrm{efficiency}&=&\frac{\textrm{number of $\pi$ tracks after $R_{\pi/K}>R_{{\rm cut}}$ }}{\textrm{number of $\pi$ tracks}}\\
K \, \textrm{misidentification probability}&=&\frac{\textrm{number of {\it K} tracks after $R_{K/\pi}<1-R_{{\rm cut}}$ }}{\textrm{number of {\it K} tracks}}\\
\pi \, \textrm{misidentification probability}&=&\frac{\textrm{number of $\pi$ tracks after $R_{\pi/K}<1-R_{{\rm cut}}$ }}{\textrm{number of $\pi$ tracks}},
\end{eqnarray}
where $R_{\rm cut}$ is the identification selection criterion and the number of $K(\pi)$ tracks is equivalent to the number of reconstructed $D^{\ast}$ decays with the $K(\pi)$ track entering the ARICH.
We determine the latter by performing an unbinned maximum likelihood fit of the $D^0$ mass distribution within the range $1.8 \, {\rm GeV}/c^2$ and $1.9 \, {\rm GeV}/c^2$.
The fit function contains a Gaussian function as the signal and a first-order polynomial function as the background contribution. 
The mean and sigma of the Gaussian are fixed by using the fitting result of the distribution before applying the $R_{K/\pi}$ or $R_{\pi/K}$ selection.
The signal yield and parameters of the first-order polynomial are determined by the fit.
Using the signal yield obtained with ($A \pm \sigma_A$) and without ($B \pm \sigma_B$) identification selection criteria applied, the efficiency and misidentification probability are determined as
$\epsilon = A/B$, and the uncertainty  $\sigma_{\epsilon}$ in efficiency is determined from 
\begin{equation}
\sigma_{\epsilon} = \frac{1}{B} \sqrt{ \epsilon^2 \sigma^2_B + \sigma^2_A - 2 \epsilon^{3/2}\sigma_A\sigma_B }.
\end{equation}
The background subtracted $R_{K/\pi}$ distributions for the $\pi$ and $K$ candidates with momenta above $0.7\, {\rm GeV}/c$ obtained from the simulated and measured data are shown in Fig.~\ref{pid}.
$K$-like particles are concentrated near 1 and $\pi$-like particles are concentrated near 0.
Figs~\ref{d0mass:mc} and \ref{d0mass:data} show $D^0$ mass distributions before and after applying the $R_{K/\pi} (R_{\pi/K})$ selection criterion for the $K (\pi)$ tracks, together with fitting curves for MC and measured data.
The $K(\pi)$ efficiency and the $\pi(K)$ misidentification probability obtained from the fitting results are summarized in Table~\ref{tab:result}.
There are differences of $3 \, \%$ between data and MC for both $K$ and $\pi$.
The corresponding receiver operating characteristic (ROC) curve is shown in Fig.~\ref{roc}.
We study the dependence of the performance on the track momentum or polar angle by dividing candidates in bins of the respective variable.
The obtained $\pi(K)$ efficiency and $K(\pi)$ misidentification probability as a function of momentum and polar angle using the value of $R_{{\rm cut}} = 0.6$ are shown in Fig.~\ref{mom} and Fig.~\ref{theta}, respectively.
These results demonstrate that the particle identification performance of the ARICH counter is good and close to expectations.
With more collected data, we will study the discrepancy between data and MC in more details to further improve the PDF and the ARICH performance.

\begin{table}[]
\centering
\caption{Comparison of overall performance between data and MC.}
\begin{tabular}{| c || c | c | c | c |}\hline
	& $K$ eff.				& $\pi$ mis.			& $\pi$ eff.			& $K$ mis.			\\ \hline \hline
Data	& $93.5 \pm 0.6 \, \%	$	& $10.9 \pm 0.9 \, \%$	& $87.5 \pm 0.9 \, \%$	& $5.6 \pm 0.3 \, \%$		\\ \hline
MC	& $96.7 \pm 0.2 \, \%$	& $7.9 \pm 0.4 \, \%$		& $91.3 \pm 0.3 \, \%	$	& $3.4 \pm 0.4 \, \%$		\\ \hline
\end{tabular}
\label{tab:result}
\end{table}

\section{Conclusion}
The ARICH counter is a novel charged particle identification device for the Belle II spectrometer designed to separate pions and kaons with momenta up to $4 \, {\rm GeV}/c$.
The Belle II experiment started its operation in 2019 and during its first period of data-taking (March - June) it collected about $5.15 \, {\rm fb}^{-1}$ of data.
Part of this data was used for the first calibration of the ARICH counter; this includes the tuning of the probability density function used for the likelihood construction, as described in this work.
Using these early Belle II data we also performed the first evaluation of the identification performance of the ARICH counter, based on pion and kaon tracks from the $D^{\ast +} \to D^0 \pi^+ (D^0 \to K^-\pi^+)$ decays.
The overall $K(\pi)$ efficiency and $\pi(K)$ misidentification probability are $93.5 \pm 0.6 \, \%$ ($87.5 \pm 0.9 \, \%$) and $10.9 \pm 0.9 \, \%$ ($5.6 \pm 0.3 \, \%$), respectively.
There is a discrepancy of about 3~\% between data and MC for both kaons and pions.
The same level of agreement between the MC and measured data is found for all momentum and polar angle regions.
These studies demonstrate that the performance of the ARICH counter during the early stage of the Belle~II operation is close to expectations.

\end{document}